\documentclass[journal]{IEEEtran}
\usepackage[T1]{fontenc}
\usepackage{graphicx}
\usepackage{subfigure}
\usepackage{cite}
\usepackage{url}
\usepackage{float}
\usepackage{graphicx}
\usepackage{textcomp}
\usepackage{setspace}
\usepackage{makecell}
\usepackage{textcomp}
\begin{document}
	
	\title{Hot Carrier Degradation in MOSFETs at Cryogenic Temperatures Down to 4.2$\,$K}
	
	\author{Yuanke Zhang, Jun Xu, Tengteng Lu, Yujing Zhang, Chao Luo and Guoping Guo
		
		\thanks{This work was supported by the National Key Research and Development Program of China (Grant No.2016YFA0301700), the National Natural Science Foundation of China (Grants No. 12034018 and 11625419), the Anhui initiative in Quantum Information Technologies (Grants No. AHY080000), and this work was partially carried out at the USTC Center for Micro and Nanoscale Research and Fabrication. \emph{(Corresponding author: Chao Luo.)}}
		\thanks{The authors are with Department of Physics, CAS Key Lab of Quantum Information, University of Science and Technology of China, Hefei 230026, Anhui, China. (e-mail: lc0121@ustc.edu.cn)}}
	
	\markboth{IEEE TRANSACTIONS ON DEVICE AND MATERIALS RELIABILITY,~Vol.~XX, No.~X, XX~2021}
	{Shell \MakeLowercase{\textit{et al.}}: Bare Demo of IEEEtran.cls for IEEE Journals}

	\maketitle
	
	\begin{abstract}
		Wide attention has been focused on cryogenic CMOS (cryo-CMOS) operation because of its promising improvement of devices' and circuits' performance and wide application prospects. However, hot carrier degradation (HCD) limits the long-term reliability of cryo-CMOS. This article investigates HCD in 0.18$\,$$\mu$m bulk CMOS at cryogenic temperature down to 4.2$\,$K. Particularly, the relationship between HCD and the current overshoot phenomenon and the influence of substrate bias on HCD are discussed. Besides, we predict the lifetime of the device at 77$\,$K and 4.2$\,$K. It is concluded that cryogenic NMOS cannot reach the ten years' commercial standard lifetime at standard $V_{DD}$. And it is predicted that the reliability requirements can be reached when $V_{DD}$$\leq $1.768$\,$V and 1.734$\,$V at 77$\,$K and 4.2$\,$K, respectively. Differently, the lifetime of PMOS is long enough even at low temperatures. 
	\end{abstract}

	
	\IEEEpeerreviewmaketitle

\section{Introduction}

\IEEEPARstart{C}{ryo-CMOS} has been widely used in neutrino physics experiments, space exploration, and has been researched for quantum computing in recent years\cite{ref1,ref2,ref51,ref62,ref63,ref65}. In order to avoid introducing excessive thermal noise during signal transmission, the interface for the quantum devices using cryo-CMOS is placed at liquid helium temperature (4.2$\,$K), providing a scalable solution  for the interface development of quantum chips\cite{ref64,ref3,ref4}. However, early research declared that HCD has a strong correlation with temperature and worsens upon cooling \cite{ref12,ref13,ref14,ref63,ref62}, which severely affects long-term cryo-CMOS operation \cite{ref5,ref12}.
\par  To date, HCD has been widely studied around room temperature (RT)\cite{ref6,ref7,ref8,ref9,ref10,ref11,ref12,ref15}, and it leads to the degradation of amplification performance, delay of digital circuits and other adverse effects \cite{ref6,ref16}.  In this paper, we investigate HCD in 0.18$\,$$\mu$m bulk MOSFETs at cryogenic temperatures down to 4.2$\,$K. The degradation mechanism is analyzed and the effect of temperature on HCD is explained physically. Particularly, the relationship between HCD and the current overshoot phenomenon and the influence of substrate bias on HCD are discussed.
\par  The goal of this work is to evaluate the reliability of 0.18$\,$$\mu$m bulk MOSFETs operating at cryogenic temperature. Therefore, we predict the cryogenic lifetime of the device, and the rated voltage that meets the reliability requirements is given. This work contributes to cryo-CMOS devices research and cryo-CMOS circuits design.

\section{Messurement Setup}
MOSFETs studied in this work were fabricated by Semiconductor Manufacturing International Corporation (SMIC) 0.18$\,$$\mu$m bulk CMOS Technology. The gate oxide thickness (Tox) is 3.6$\,$nm and $V_{DD}$$\,$=$\,$1.8$\,$V. The sample chips were bonded to the chip-carriers with aluminum wires, as shown in Fig. 1(a). The cryogenic measurements were performed in liquid nitrogen Dewar, liquid helium Dewar, and a 1.2$\,$K refrigerator cryostat. All the MOSFETs' electrical characteristics measurements and stress were performed by a Keysight B1500A semiconductor device analyzer and measured by the four-wire method to remove the influence of wire resistance. The stress was periodically interrupted to measure the device parameters.
\par To analyze the degradation of the device characteristics, transfer characteristics in both linear region ($V_{DS}$$\,$=$\,$50$\,$mV) and saturation region ($V_{DS}$$\,$=$\,$1.8$\,$V) were measured with the source and substrate terminal grounded. The maximum transconductance ($G_{mmax}$) was obtained from linear region transfer characteristics, and the threshold voltage $V_{TH}$ was extracted at the threshold current $I_{TH}$$\,$=$\,$$1.0\times10^{-8}$ $W$/$L$$\,$(A). In the measurements of PMOS, the above values were replaced by the opposite number. 

\section{Results and Discussions}

\subsection{Low temperature characterization}

Low temperature operation can improve the performance of MOSFETs, as shown in Fig. 1(b), Table \ref{tab1}, and Table \ref{tab2}. At low temperatures, the off-state current ($I_{OFF}$, $V_{DS}$$\,$=$\,$1.8$\,$V and $V_{GS}$$\,$=$\,$0$\,$V) decreases and the saturation current ($I_{DSAT}$, $V_{DS}$$\,$=$\,$$V_{GS}$$\,$=$\,$1.8$\,$V) increases, which greatly improves the ON/OFF ratio. $V_{TH}$ increases at low temperatures, which can be attributed to the decrease of intrinsic carrier concentration and the increase of Fermi potential.
\par Due to the longer band-to-band tunneling (BTBT) distance at low temperatures\cite{ref29}, the gate-induced drain leakage (GIDL) effect is ameliorated obviously [Fig. 1(b)]. The subthreshold swing ($SS$, obtained at $|V_{DS}|$=$50$$\,$$mV$) is given by $SS=ln(10)nK_{B}T/q$, where $n$, $K_{B}$, $T$, and $q$ is subthreshold slope factor, Boltzmann constant, Kelvin temperature, and the electron charge, respectively. Therefore, as the temperature decreases, $SS$ decreases significantly, resulting in faster switching speeds of the device. The low-field mobility ($\mu_{0}$) is extracted from the linear region transfer characteristics and can be expressed as\cite{ref29}: 
\begin{equation}
\mu_{0}=\frac{G_{mmax} L}{C_{OX}V_{DS}W}
\end{equation}
where $C_{OX}$ is the gate oxide capacitance and the given $|V_{DS}|$ is $50$$\,$$mV$ for our MOSFETs. At low temperatures, the reduced lattice scattering leads to larger $\mu_{0}$ and larger $G_{m}$, as shown in Table \ref{tab1} and Fig. 1(c), respectively. Differently, $\mu_{0}$ and $G_{m}$ of PMOS are the largest at 77$\,$K, as shown in Table \ref{tab2} and Fig. 1(d). For PMOS, due to the difference between the work function of the poly gate and the substrate, a light boron implant in the channel is required to adjust $V_{TH}$, which leads to the formation of the buried-channel. Due to the freeze-out of the implant, the peak of buried-channel mobility is observed around 80$\,$K \cite{ref26,ref27,ref28}. Therefore, the maximum $\mu_{0}$ and $G_{m}$ are observed at 77$\,$K.

\begin{figure}[ht]
	\centering{
		{\includegraphics[width=\linewidth]{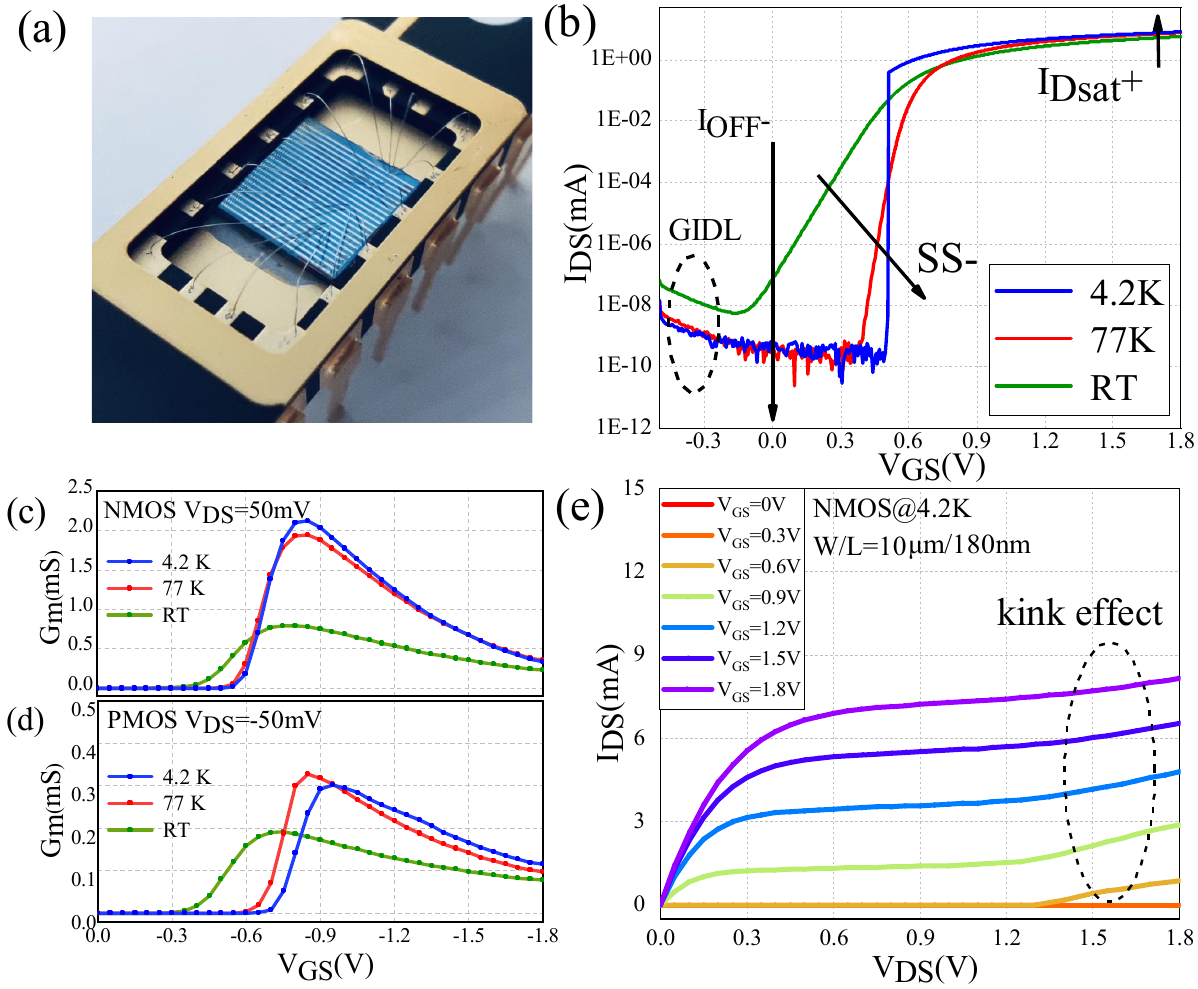}}
	}
	\caption{(a) Sample chip, wire-bonded to a chip carrier with Al-wire bonds. (b) $I_{DS}$ versus $V_{GS}$ at RT, 77$\,$K, and 4.2$\,$K with $V_{DS}$$\,$=$\,$1.8$\,$V in NMOS. $G_{m}$ versus $V_{GS}$ at RT, 77$\,$K, and 4.2$\,$K in NMOS (c) and PMOS (d). (e) Output characteristics in W/L$\,$=$\,$10$\,$$\mu$m/180$\,$nm NMOS at 4.2$\,$K. The kink effect can be observed.
	}
\end{figure}

The kink effect is observed in the output curve at 4.2$\,$K [Fig. 1(e)], which is attributed to the decrease of $V_{TH}$ caused by the accumulation of holes in the substrate\cite{ref30}. In addition, the kink effect can also cause an abnormally steep rise of $I_{DS}$ at 4.2$\,$K, as shown in Fig. 1(b). When $V_{GS}$ overcomes $V_{TH}$, the channel is formed and the holes generated by impact ionization accumulate in the freeze-out substrate. Hence $V_{TH}$ decreases and $I_{DS}$ increases (i.e., the kink effect). The increased $I_{DS}$ generates more holes by impact ionization and repeats the above whole process. Therefore, the increase of $I_{DS}$ is an avalanche process and reaches the final value when the channel is in strong inversion\cite{ref30}, thus resulting in the abnormally steep rise of $I_{DS}$ from the off-state to the strong inversion. This phenomenon is only observed when $V_{DS}$ is large enough ($V_{DS}$$\geq$1.4$\,$V in 10$\,$$\mu$m/180$\,$nm NMOS, i.e., after the kink), and it is not obviously observed in PMOS. 
 
\par  Coulomb blockade oscillations are observed in small-size devices at deep-cryogenic temperatures, as shown in Fig. 2. The quasi-periodic Coulomb diamond suggests the existence of quantum dots (QD) in the channel\cite{ref31}. At deep-cryogenic temperature and under low $V_{DS}$ and low $V_{GS}$ bias, $I_{DS}$ mainly depends on the electron transport between  source, drain, and the quantum dot. This indicates that 0.18$\,$$\mu$m MOSFETs can be used to construct quantum circuits or quantum-classical hybrid circuits.

\begin{table}
	\caption{DC CHARACTERISTICS OF NMOSFETS (W/L$\,$=$\,$10$\,$$\mu m$/180$\,$$nm$) AT 300$\,$K ,77$\,$K, AND 4.2$\,$K}
	\centering
	\label{table1}
	\setlength{\tabcolsep}{6pt}
	\begin{spacing}{1}
		\begin{tabular}{|m{2cm}<{\centering}|m{1.7cm}<{\centering}|m{1.7cm}<{\centering}|m{1.7cm}<{\centering}|}
			\hline
			Temperature& 
			300$\,$K& 
			77$\,$K&
			4.2$\,$K\\
			\hline
			$V_{TH} (V)$& 
			0.356& 
			0.557&
			0.591 \\
			\hline
			$SS (mV/dec)$& 
			85.05& 
			28.44& 
			12.54 \\
			\hline
			$I_{DSAT} (mA)$& 
			5.87& 
			7.80& 
			8.45 \\
			\hline
			$I_{OFF} (A)$& 
			$7.6\times10^{-11}$& 
			$4.8\times10^{-13}$& 
			$4.6\times10^{-13}$ \\
			\hline
			$\mu_{0} (cm^{2}/Vs)$& 
			$3.24\times10^{2}$&
			$8.95\times10^{2}$& 
			$8.68\times10^{2}$ \\
			\hline
		\end{tabular}
	\end{spacing}
	\label{tab1}
\end{table}

\begin{table}
	\caption{DC CHARACTERISTICS OF PMOSFETS (W/L$\,$=$\,$10$\,$$\mu m$/180$\,$$nm$) AT 300$\,$K ,77$\,$K, AND 4.2$\,$K}
	\centering
	\label{table2}
	\setlength{\tabcolsep}{6pt}
	\begin{spacing}{1}
		\begin{tabular}{|m{2cm}<{\centering}|m{1.7cm}<{\centering}|m{1.7cm}<{\centering}|m{1.7cm}<{\centering}|}
			\hline
			Temperature& 
			300$\,$K& 
			77$\,$K&
			4.2$\,$K\\
			\hline
			$V_{TH} (V)$& 
			0.401& 
			0.659&
			0.738 \\
			\hline
			$SS (mV/dec)$& 
			86.41& 
			31.85& 
			15.15 \\
			\hline
			$I_{DSAT} (mA)$& 
			2.20& 
			2.49& 
			2.52 \\
			\hline
			$I_{OFF} (A)$& 
			$4.0\times10^{-11}$& 
			$2.1\times10^{-13}$& 
			$2.0\times10^{-13}$ \\
			\hline
			$\mu_{0} (cm^{2}/Vs)$& 
			$0.78\times10^{2}$&
			$1.34\times10^{2}$& 
			$1.23\times10^{2}$ \\
			\hline
		\end{tabular}
	\end{spacing}
	\label{tab2}
\end{table}

\begin{figure}[!t]
	\centering{
		{\includegraphics[width=3in]{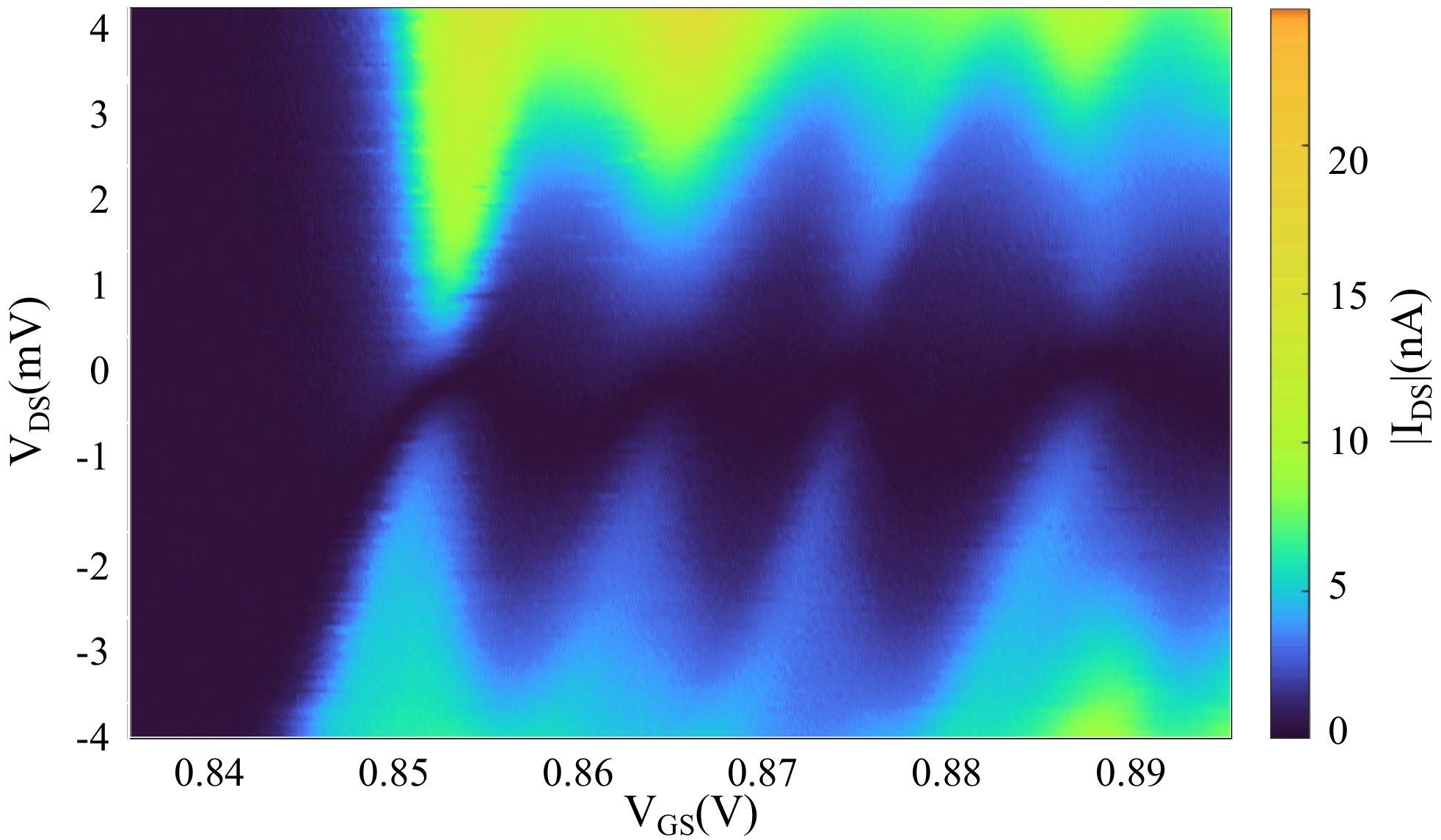}}
	}
	\caption{Coulomb blockade in 180$\,$nm/180$\,$nm NMOS at 1.2$\,$K. }
\end{figure}

\subsection{HCD at low temperatures}


It is necessary to investigate the worst-case condition of 0.18$\,$$\mu$m MOSFETs to determine the limit of device lifetime, hence we carried out measurements on NMOS and PMOS in the case of $V_{Gstress}$$\,$=$\,$$V_{Dstress}$ (channel-hot electron degradation mode) and $V_{Gstress}$@$I_{submax}$ (drain-avalanche hot carrier degradation mode, $V_{Gstress}$$\,$$\approx$$\,$1/2$V_{Dstress}$) at 300$\,$K, 77$\,$K, and 4.2$\,$K, respectively. Under a given $V_{DS}$, the substrate current ($I_{sub}$) shows a bell-shaped behavior with the change of $V_{GS}$ and $V_{Gstress}$@$I_{submax}$ is the gate voltage corresponding to the maximum $I_{sub}$. The results show that the worst-case of NMOS is $V_{Gstress}$@$I_{submax}$ and the worst-case of PMOS is $V_{Gstress}$$\,$=$\,$$V_{Dstress}$ at each measurement temperature. Given the difference of worst-case stress, NMOS and PMOS were tested under their respective worst-case condition in the following measurements. Besides, since the degradation process is inversely proportional to the length of the device\cite{ref25}, the measurements were taken on short channel devices to accelerate the HCD process.
 
\begin{figure}[ht]
	\centering{
		\includegraphics[width=\linewidth]{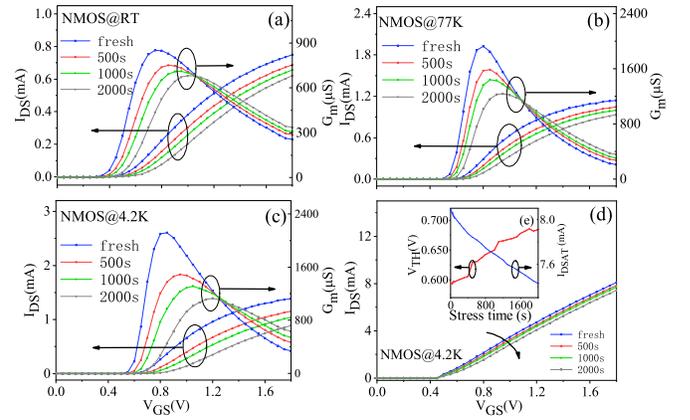}
	}
	\caption{At $V_{Dstress}$$\,$=$\,$2.8$\,$V, $V_{Gstress}$@$I_{submax}$ in 10$\,$$\mu$m/180$\,$nm NMOS, stress time$\,$=$\,$0$\,$s/500 $\,$s/1000$\,$s/2000$\,$s: linear region transfer characteristic and $G_{m}$ versus $V_{GS}$ at RT (a), 77$\,$K (b) and 4.2$\,$K (c). (d) Saturation region ($V_{DS}$$\,$=$\,$1.8$\,$V) transfer characteristic at 4.2$\,$K. Inset: $V_{TH}$ an $I_{DSAT}$ versus stress time at 4.2$\,$K.}
	\label{fig4}  
\end{figure}

Fig. \ref{fig4}(a)-(c) show the linear-region transfer characteristic degradation and transconductance degradation under different stress time at RT, 77$\,$K, and 4.2$\,$K, respectively. The peak of transconductance shifts to more positive $V_{GS}$ with increasing stress time. With the decrease of temperature, the transconductance curves under different stress time become more dispersed, indicating that HCD is more severe at cryogenic temperatures, which is consistent with the conclusion in \cite{ref62,ref63}. Fig. 3(d) shows the saturation-region transfer characteristic degradation. As shown in Fig. 3(e), $V_{TH}$ and $I_{DSAT}$ is positively and negatively shifted, respectively. The shift of $V_{TH}$ indicates the accumulation of negative charges in the gate oxide, which results from hot carriers trapped into the oxide or charged interface created at the oxide interface. Degradation of $G_{mmax}$ is used to analyze the HCD effect in this article, and the lifetime is defined as 10$\%$ $G_{mmax}$ degradation. 

\begin{figure}[ht]
	\centering{
		\includegraphics[width=\linewidth]{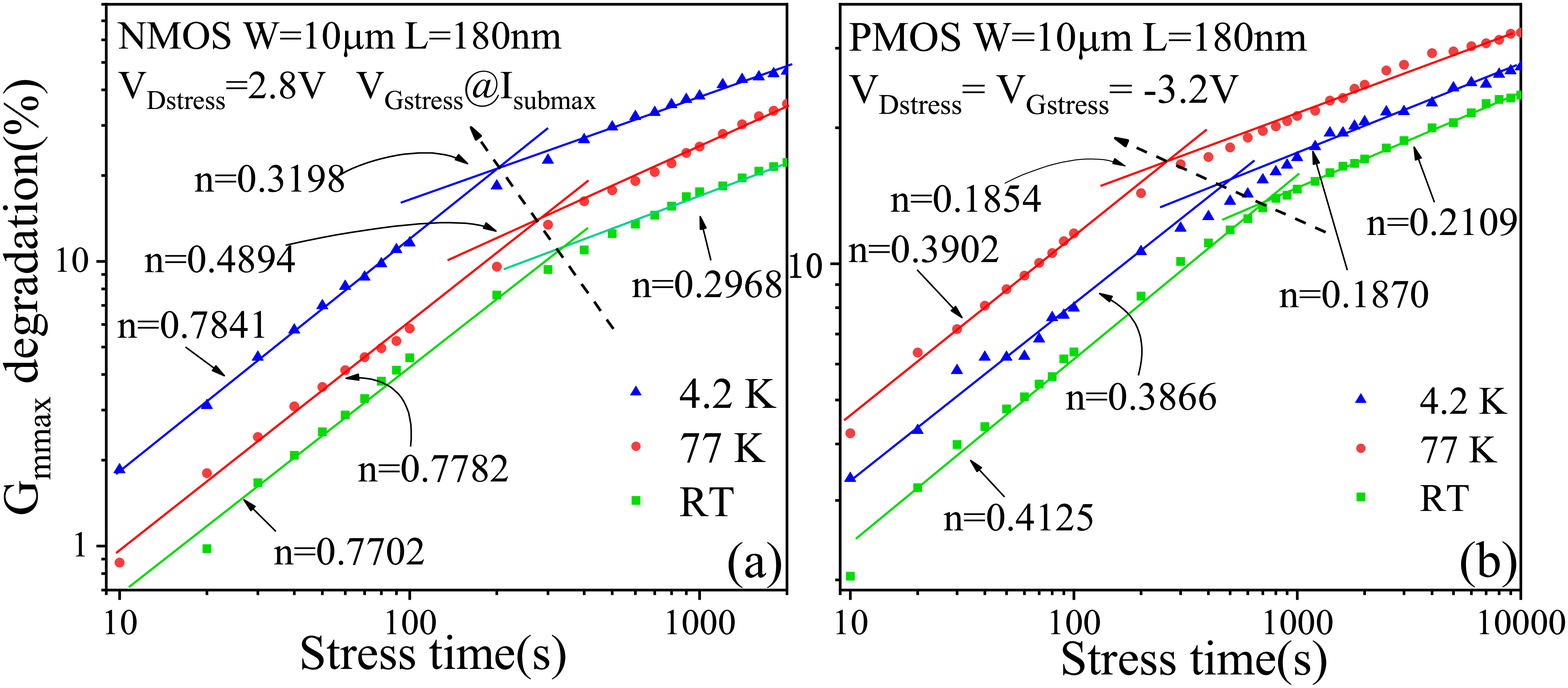}
	}
	\caption{ $G_{mmax}$ degradation versus stress time in 10$\,$$\mu$m/180$\,$nm NMOS (a) and PMOS (b) at RT, 77$\,$K, and 4.2$\,$K, plotted in a log-log scale.}
		\label{fig5} 
\end{figure}

Degradation of device parameters complies with the power-law function of time \cite{ref6,ref10,ref15}:

\begin{equation}
G_{mmax}(I_{DSAT},V_{TH})\ degradation \propto t^n
\end{equation}

\noindent where the parameter $n$ is the time power-law exponent and $t$ is the stress time. If HCD is due to interface state damage ($N_{it}$ degradation), $n$ is in the range of 0.5 to 1 and hot-carrier injecting into the gate oxide ($N_{ot}$ degradation) leads to a smaller $n$ from 0.1 to 0.3\cite{ref6,ref12,ref22}. Fig. \ref{fig5}(a) shows the time power-law exponent of NMOS at RT, 77$\,$K, and 4.2$\,$K. The values of power-exponent $n$ are similar at different temperatures and it indicates that the degradation mechanism is similar at different temperatures: the initial degradation is mainly caused by $N_{it}$ degradation and it changes into $N_{ot}$ degradation as the stress time increases. A time-varying power-law exponent can be used to characterize the HCD process at not only RT but also cryogenic temperatures.
\par Meanwhile, the relationship between the slope change point and the temperature is noticeable. The interface trap charges are generated and accumulated near the drain side. At cryogenic temperatures, the carrier mobility and the energy of hot carriers increase, hence hot carriers can accumulate enough energy closer to the source to form interface traps, resulting in a larger area of interface trap charge accumulation. Therefore, the larger area of interface traps at cryogenic temperatures leads to more severe $N_{it}$ degradation, and the slope change point shifts to more positive $G_{mmax}$ degradation [dotted line in Fig. \ref{fig5} (a)].
\par Similar phenomena have also been observed in PMOS, as shown in Fig. \ref{fig5}(b). The initial mechanism is the mixing mechanism of $N_{it}$ degradation and $N_{ot}$ degradation. Because the formation of $N_{it}$ saturates with increasing stress time, $N_{ot}$ degradation is the HCD mechanism for long-term stress time. In addition, due to the higher mobility of PMOS at 77$\,$K (Table \ref{tab2}), HCD at 77$\,$K is more severe than that at 4.2$\,$K.

\subsection{HCD and the current overshoot}
\par
\begin{figure}[!t]
	\centering{
		\includegraphics[width=\linewidth]{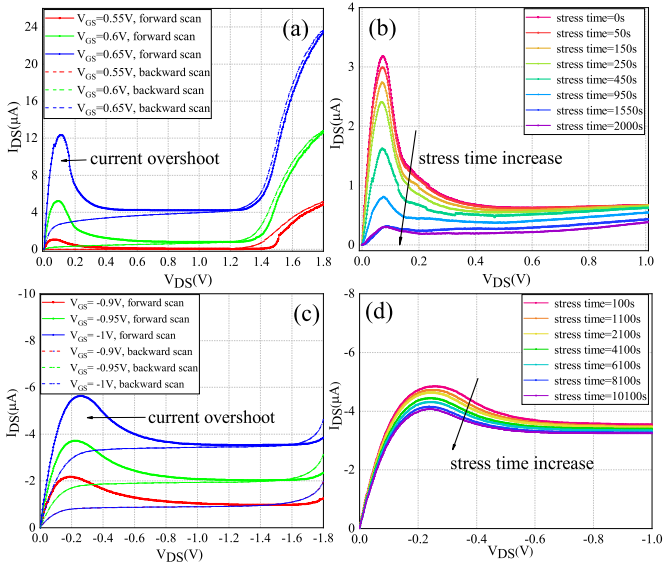}
	}
	\caption{ (a) Forward scans (solid lines) and backward scans (dashed lines) of I-V characteristics in 10$\,$$\mu$m/10$\,$$\mu$m NMOS at 4.2$\,$K. $V_{GS}$$\,$=$\,$0.55$\,$V, 0.6$\,$V, and 0.65$\,$V, $V_{DS}$ step is 5$\,$mV. (b) The current overshoot phenomenon under different stress time, $V_{GS}$$\,$=$\,$0.6$\,$V and $V_{DS}$ step is 2$\,$mV. $V_{Dstress}$$\,$=$\,$4$\,$V and $V_{Gstress}$@$I_{submax}$. (c) Forward scans (solid lines) and backward scans (dashed lines) of I-V characteristics in 10$\,$$\mu$m/10$\,$$\mu$m PMOS at 4.2$\,$K. $V_{GS}$$\,$=$\,$-0.9$\,$V, -0.95$\,$V, and -1$\,$V, $V_{DS}$ step is -5$\,$mV. (b) The current overshoot phenomenon under different stress time, $V_{GS}$$\,$=$\,$-1$\,$V and $V_{DS}$ step is -2$\,$mV. $V_{Dstress}$$\,$=$V_{Gstress}$$\,$=$\,$-5$\,$V. }
		\label{fig6} 
\end{figure}

 The current overshoot phenomenon at cryogenic temperatures is widely reported in large-size transistors (from 96$\,$$\mu$m/12$\,$$\mu$m to 10$\,$$\mu$m/10$\,$$\mu$m)\cite{ref62,ref33,ref71,ref72}, but it is not reported in small size transistors. This is consistent with our measurement results: as shown in Fig. \ref{fig6}(a), the current overshoot phenomenon is observed in 10$\,$$\mu$m/10$\,$$\mu$m NMOS at 4.2$\,$K and not observed in 10$\,$$\mu$m/180$\,$nm NMOS [Fig. 1(e)]. 

\par The current overshoot phenomenon can be attributed to the trapped charges in the Si-SiO$_{2}$ interface and the positive charges in the interface traps can lead to the decrease of $V_{TH}$ and the increase of $I_{DS}$. When $V_{DS}$ increases further, $I_{DS}$ decreases, which can be attributed to two reasons: (i) The increase of $V_{DS}$ leads to the channel pinch-off, forming the depletion region at the drain side and shortening the effective channel length, so that some interface trap charges do not affect the threshold voltage. (ii) The positive charges in the interface traps can be emptied, and the emptying process is a function of time\cite{ref33}. Therefore, as the forward scanning proceeds, the positive charges in the interface gradually decrease, resulting in the decrease of $I_{DS}$. The backward scan is performed continuously after the forward scan, so most of the positive charges in the interface traps have been emptied. Therefore, the current overshoot phenomenon does not appear during the backward scan.

\par As shown in Fig. \ref{fig6}(b), the peak value of the current overshoot decreases obviously after hot carrier injection. The injection of hot electrons neutralizes the effect of the positive charges in the interface traps. Therefore, with the increase of stress time, overshoot becomes weaker, and it can be predicted that the current overshoot phenomenon will disappear under sufficient stress time. Similar phenomenons are also observed in PMOS, as shown in Fig. \ref{fig6}(c) and (d). The relationship between the hot carrier effect and the current overshoot phenomenon confirms that the overshoot is caused by the interface states. The self-heating effect is another explanation of the current overshoot\cite{ref32}. In our measurements, the overshoot phenomenon in NMOS is observed at medium $V_{GS}$ ($V_{GS}$$\,$$\approx$$\,$0.5$\,$V-1$\,$V), but it is not observed at large $V_{GS}$ (i.e., large $I_{DS}$ condition). Therefore, this explanation is not supported in this article.

\subsection{The influence of substrate bias}

\begin{figure}[h]
	\centering
	\includegraphics[width=\linewidth]{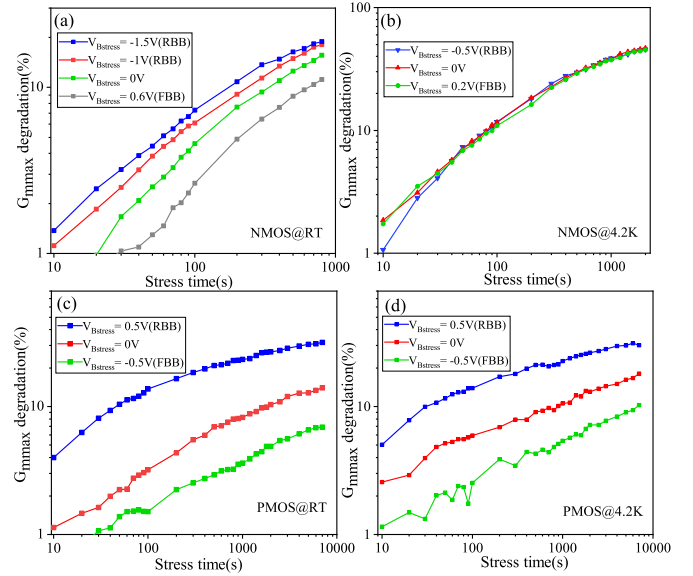}
	\caption{$G_{mmax}$ degradation under various $V_{Bstress}$ at $V_{Dstress}$$\,$=$\,$2.8$\,$V, $V_{Gstress}$@$I_{submax}$  in 10$\,$$\mu$m/180$\,$nm NMOS at RT (a) and 4.2$\,$K (b). $G_{mmax}$ degradation under various $V_{Bstress}$ at $V_{Dstress}$$\,$=$\,$$V_{Gstress}$$\,$=$\,$-3.2$\,$V in 10$\,$$\mu$m/200$\,$nm PMOS at RT (c) and 4.2$\,$K (d).}
	\label{fig7} 
\end{figure}

The effect of substrate bias on the HCD process has only been studied around room temperature, and no common conclusion is reached\cite{ref17,ref19}. Hence, we investigate HCD with substrate bias at RT and 4.2$\,$K. As shown in Fig. 6(a), at RT, forward substrate bias (FBB) can significantly reduce degradation, and the degradation becomes more severe as the substrate bias becomes more negative in NMOS. This is consistent with \cite{ref19} but different from \cite{ref17}. In the pinch-off region, the average value of the transverse (i.e., parallel to the channel) electric field $\overline{\mathcal{E}_{x}}$ can be expressed as:

\begin{equation}
\overline{\mathcal{E}_{x}} \approx \frac{V_{DS}-V_{DSAT}}{l}
\end{equation}

\noindent where $l$ is the length of the pinch-off region, $V_{DSAT}$ is the saturation drain voltage, and $V_{DSAT}$\,$\approx$\,$V_{GS}-V_{TH}$. With FBB, $V_{TH}$ decreases and $V_{DSAT}$ increases, resulting in the decrease of the transverse electric field. The effective longitudinal (i.e., perpendicular to the channel) electric field $\mathcal{E}_{y}$ can be expressed as:

\begin{equation}
\mathcal{E}_{y}=\frac{1}{\varepsilon_{s}}\left(\frac{Q_{i}}{2}+Q_{b}\right)
\end{equation}

\noindent where $\varepsilon_{s}$ is the dielectric constant, $Q_{i}$ is the inversion layer charge per unit area, and $Q_{b}$ is the depletion region charge per unit area. Under FBB, the depletion region becomes thinner and $Q_{b}$ decreases, so that $\mathcal{E}_{y}$ decreases. Hence, in the pinch-off region ($Q_{i}\approx0$), both the transverse and the  longitudinal electric field decrease under FBB. The decrease of transverse electric field reduces the impact of carriers and the generation of hot carriers, and the decrease of longitudinal electric field reduces the carriers injecting into the gate oxide. Therefore, the reliability of devices is improved with FBB and degraded with reverse substrate bias (RBB).





\par The HCD process is almost the same under various $V_{Bstress}$ at 4.2$\,$K, as shown in Fig. \ref{fig7}(b). This can be attributed to the freeze-out of the substrate. At 4.2$\,$K, the holes produced by carriers impact ionization accumulate in the freeze-out substrate, forming a positive substrate fixed potential and leading to the kink effect [Fig. 1(d)]. Therefore, the kink effect indicates the formation of floating substrate potential and the loss of control of the substrate potential\cite{ref27}, thus resulting in the HCD process not being affected by $V_{Bstress}$. Therefore, FBB is ineffective in extending the worst-case lifetime of NMOS at 4.2$\,$K.

\par Differently, FBB can enhance the lifetime of PMOS at both RT and 4.2$\,$K, as shown in Fig. \ref{fig7}(c) and (d). We performed measurements under the worst-case condition of PMOS (i.e., $V_{Gstress}$$\,$=$\,$$V_{Dstress}$), but the maximum $I_{sub}$ is near $V_{GS}$$\,$=$\,$$1/2V_{DS}$. The small $I_{sub}$ at $V_{GS}$$\,$=$\,$$V_{DS}$ can not provide enough carriers to the freeze-out substrate to form a floating substrate potential, which means that the substrate potential is still controllable. Therefore, FBB can improve the worst-case lifetime of PMOS at both RT and cryogenic temperatures. 

\subsection{Lifetime prediction}

The low-temperature lifetime prediction of deep sub-micrometer MOSFETs has been carried out at 77$\,$K\cite{ref5}, and we extended this work to 4.2$\,$K. Considering the influence of $V_{DS}$, the HCD power-law relation can be written as \cite{ref24,ref25}:
\begin{equation}
Parameter\ degradation\% = \alpha t^nexp(\beta/V_{DS})
\end{equation}
\noindent where $\alpha$ is a positive constant, $\beta$ is a negative constant, and $n$ is the time power-law exponent above. When the characteristic degradation percentage of a certain parameter is taken as the failure criterion, the lifetime ($\tau$) can be expressed as\cite{ref5,ref6}:
\begin{equation}
\tau = Aexp(B/V_{DS})
\end{equation}
\noindent where $A$ is a constant and $B$$\,$=$\,$$-\beta$/$n$. Because the change of $n$ all occurs after 10\% $G_{mmax}$ degradation, $B$ is considered as a constant in this paper. If the definition of lifetime is different, such as 20\% $G_{mmax}$ degradation, then the change of $n$ should be considered in the lifetime prediction. In addition, according to the conclusion in subsection B, with the further increase of temperature ($T\textgreater$300$\,$K), $n$ will change before 10\% $G_{mmax}$ degradation [see Fig. \ref{fig5}(a)]. Therefore, Eq. (6) needs to be used carefully at high temperatures.

\begin{figure}[ht]
	\centering{
		\includegraphics[width=\linewidth]{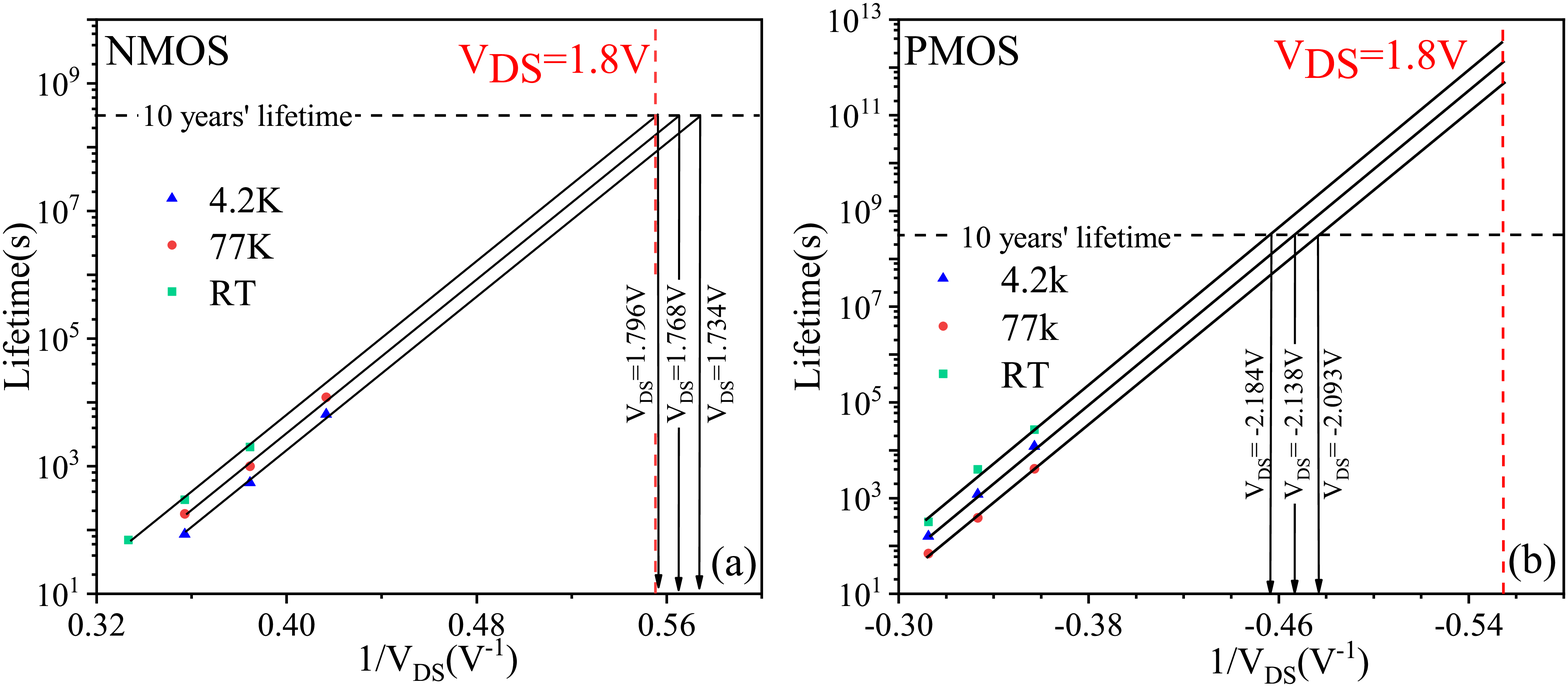}
	}
	\caption{ Lifetime prediction of 10$\,$$\mu$m/180$\,$nm NMOS (a) and PMOS (b) at RT, 77$\,$K, and 4.2$\,$K. The rated $V_{DD}$ of ten years' lifetime: 1.796$\,$V, 1.768$\,$V, and 1.734$\,$V for NMOS and 2.184$\,$V, 2.093$\,$V, and 2.138$\,$V for PMOS at RT, 77$\,$K, and 4.2$\,$K, respectively.}
	\label{fig9} 
\end{figure}

\par For NMOS, the lifetime is tested at the stress voltages: $V_{Dstress}$$\,$=$\,$2.8$\,$V, 2.6$\,$V, and 2.4$\,$V (2.4$\,$V was replaced by 3.0$\,$V at RT), $V_{Gstress}$@$I_{submax}$. For PMOS, the lifetime is tested at $V_{Dstress}$$\,$=$\,$$V_{Gstress}$$\,$=$\,$-3.2$\,$V, -3$\,$V, and -2.8$\,$V. According to Eq. (6), experimental data is extrapolated, as shown in Fig. \ref{fig9}(a) and (b). The lifetime prediction of some common DC voltage is also shown in Table \ref{tab3}. It shows that NMOS cannot meet the requirement of ten years' lifetime under standard $V_{DD}$, which affects cryo-CMOS long-term reliability.
\par Due to the steeper $SS$ and the ameliorated ON/OFF ratio at cryogenic temperatures, slightly reducing $V_{DD}$ provides a solution to improve the lifetime without affecting the circuits' performance, and also benefits the low power consumption design of cryo-CMOS. We calculated the rated $V_{DD}$ for ten years' lifetime, as shown in  Fig. \ref{fig9}(a). At 77$\,$K and 4.2$\,$K, $V_{DD}$ should be less than 1.768$\,$V and 1.734$\,$V to meet cryo-CMOS reliability requirements, respectively. Fortunately, the lifetime of PMOS at each temperature is much longer than ten years, as shown in Fig. \ref{fig9}(b). The holes have shorter mean free path, greater effective mass, and higher Si-SiO$_{2}$ potential energy barriers than the electrons, hence the lifetime of PMOS is much longer than that of NMOS.

\begin{table}[H]
	\centering
	\caption{LIFETIME PREDICTION (YEARS)}
	
	\begin{tabular}{cccc}
		\hline
		DC voltage	& 0.9$V_{DD}$ & 1.1$V_{DD}$ &Standard $V_{DD}$$\,$=$\,$1.8$\,$V  \\
		\hline
		NMOS@RT	&  654&0.283 & 9.24  \\
		
		NMOS@77K&  349&0.150 &  4.90 \\
		
		NMOS@4.2K&  161& 0.07 &2.27  \\
		
		PMOS@RT	& 2.61e+7 & 824 & 8.72e+4 \\
		
		PMOS@77K& 3.62e+6 & 114 &  1.22e+4\\
		
		PMOS@4.2K&  1.06e+7&  336& 3.56e+4 \\
		
		\hline
		\label{tab3}
	\end{tabular}
	
\end{table}

\section{Conclusion}

A study of HCD at cryogenic temperature down to 4.2$\,$K is presented in this article. Particularly, the relationship between HCD and the current overshoot phenomenon and the influence of substrate bias on HCD are discussed. Besides, the lifetime of the device is predicted. For NMOS, HCD becomes more severe and the lifetime cannot reach ten years' commercial standard lifetime at cryogenic temperatures. By extrapolating the measurement data, it is concluded that reducing $V_{DD}$ to 1.768$\, $V and 1.734$\, $V can reach the rated lifetime at 77$\, $K and 4.2$\, $K, respectively. Fortunately, PMOS has sufficient reliability at each temperature.

\appendices

\ifCLASSOPTIONcaptionsoff
  \newpage
\fi

\end{document}